# Tevatron Beam Position Monitor Upgrade*

**G. Annala[a], B. Banerjee[a], B. Barker[a], T. Boes[a], M. Bowden[a], C. Briegel[a], G. Cancelo[a], G. Duerling[a], B. Forster[a], S. Foulkes[a], B. Haynes[a], B. Hendricks[a], T. Kasza[a], R. Kutschke[a], R. Mahlum[a], M. Martens[a], M. Olson[a], V. Pavlicek[a], L. Piccoli[a], P. Prieto[a†], J. Steimel[a], K. Treptow[a], M. Votava[a], D. Voy[a], M. Wendt[a], S. Wolbers[a], D. Zhang[a]**

[a] *Fermi National Accelerator Laboratory*
*P.O. Box 500 , Batavia, IL, 60510, U.S.A.*
*E-mail*: Prieto@FNAL.gov

ABSTRACT: This paper describes the development of a digital-based Beam Position System which was designed, developed, and adapted for the Tevatron during Collider Run II.

* Work supported by Fermi Research Alliance, LLC under Contract No. De-AC02-07CH11359 with the United States Department of Energy.

† Corresponding author

## Contents

## 1. Introduction

For a successful Run II to occur at Fermilab, multiple diagnostic systems in the various accelerators required upgrades. This article deals with the upgrade of the Tevatron (TeV) BPM system. Additional material in this article came from the references included at the end. The original Tevatron BPM system was designed when the TeV was being commissioned in 1981. This was an analog-based system designed by a team lead by Robert Shafer and Bob Webber. Although the system was adequate for the time and provided excellent and reliable service over the years an upgraded system was required to meet the requirements of the Tevatron during Run II.

The digital-based BPM system that replaced the old analog system is based on a design implemented by Jim Sebek at SLAC, who worked closely with Echotek Corporation in adapting an eight-channel digital receiver (DDC) board produced for radar applications. At Fermilab the DDC board became the basis of all the BPM upgrades implemented on all accelerators (Recycler, TeV, Main Injector and transfer lines) starting in 2003. The first BPM system developed under this new architecture was the Recycler BPM system consisting of 220 BPMs, which combined an analog signal conditioning hardware with the VME64X-based down converter to produce turn-by-turn, and closed orbit beam measurements. Once this technology was proven successful in the Recycler Ring it was then implemented in the Tevatron (240 BPMs), followed by the Main Injector BPM system (210 BPM's) and beam transport lines.

The advantages this digital-based BPM system brings to the user are significant. First, each beam signal can be measured and displayed in real time in the control room. Second, the processing modes can be implemented by changing the signal processing filter parameters and then downloading them into the DDC boards without having to re-design the hardware which is time consuming. Thirdly, various operational modes can be switched in real-time triggered by machine timing events and/or machine states.

## 2. Tevatron BPM Upgrade

### 2.1 Tevatron BPM Requirements

In this section the list of requirements of the BPM system is included. The acronyms used in the tables are TCLK (Tevatron Clock), ACNET (accelerator controls network), FTP (fast time plot).

Note concerning the fourth column of table 1. Even though the Tevatron BPM requirements document states the position and accuracy resolution to be the listed values in the fourth column, what was really meant here is a required resolution should be the specified values, not also accurate to the values listed.

**Table 1.** Tevatron Measurement Requirements.

| Measurement Purpose | Beam Structure | Data Acquisition Type | Position Accuracy and Resolution |
|---|---|---|---|
| Proton closed orbit during a store. | 36 X 36. | Manual.<br>Buffered on TCLK.<br>ACNET variable.<br>FTP variable. | Position Resolution of 0.007 mm. |
| Proton single turn for injection tune up. | Proton un-coalesced. | Single turn, triggered on TCLK. | Position Resolution of 0.05 mm. |
| PBAR closed orbit during a store. | 36X36. | Manual.<br>Buffered on TCLK.<br>ACNET variable.<br>FTP variable. | Position resolution of 0.05mm. |
| Proton closed orbit during ramp and LB squeeze. | 36X36.<br>Proton coalesced.<br>Proton un-coalesced. | Buffered on TCLK.<br>ACNET variable.<br>FTP variable. | Position resolution of 0.05mm. |
| Proton single turn for injection commissioning. | Proton un-coalesced. | Single turn, triggered on TCLK. | Position resolution of 0.1mm. |
| Proton closed orbit for injection commissioning | Proton un-coalesced. | Buffered on TCLK. | Position resolution of 0.05mm. |
| Proton single turn for injection tune up. | Proton un-coalesced. | Single turn, triggered on TCLK. | Position resolution of 0.05mm. |
| Proton closed orbit for injection tune up. | Proton un-coalesced. | Buffered on TCLK. | Position resolution of 0.02mm. |
| Closed orbit circular buffer. | 36X36.<br>Proton un-coalesced.<br>Proton coalesced.<br>Pbar coalesced. | Circular buffer halted on Tevatron abort. | Position resolution of 0.007mm. |
| Aperture scans. | Proton un-coalesced.<br>Proton coalesced. | Manual.<br>Buffered on TCLK.<br>ACNET variable.<br>FTP variable. | Position resolution of 0.007mm. |
| Lattice measurements. | Proton un-coalesced.<br>Proton coalesced. | Manual.<br>Buffered on TCLK.<br>ACNET variable.<br>FTP variable. | Position resolution of 0.007mm. |
| Lattice and coupling measurements. | Proton un-coalesced.<br>Proton coalesced. | TBT buffer. | Position resolution of 0.007mm. |

**Table 2.** Range of Intensities and bunch lengths in Run II.

| | Particles/Bunch | Number of Bunches | Bunch Length<br>(3σ value in nsec.)<br>RF 53.1048 MHz |
|---|---|---|---|
| Un-coalesced Protons | 3e9 to 30e9 | 30 | 3.5 to 10 |
| Coalesced Protons | 30e9 to 350e9 | 1 to 36 | 4.5 to 10 |
| Coalesced Antiprotons | 3e9 to 150e9 | 1 to 36 | 4.5 to 10 |

Tevatron beam can be coalesced or un-coalesced for protons as well as antiprotons. The un-coalesced mode is a train of 20 to 30 consecutive bunches spaced at 18.83 nanoseconds injected into the Tevatron.

Coalesced beam is made by coalescing a train of protons into a single bunch in the Main Injector before transferring it into the Tevatron. This process is done repeatedly to create three trains of 12 coalesced bunches each, both of protons and antiprotons. The bunches in a single train are separated by 21 RF buckets (395.44 nanoseconds), where an RF bunch is 18.83 nanoseconds and the trains are separated by 140 RF buckets (2.636 μseconds) [1].

### 2.2 BPM Pickups

The BPM pickups in the Tevatron ring are part of the superconducting quadrupole assemblies and were not modified as part of the BPM upgrade. Each BPM is a pair of 50 Ω striplines, 18 cm long, each subtending 110 degrees of arc with a circular aperture of 7.0 cm diameter. Each BPM measures either the vertical or the horizontal coordinate, and there are approximately 240 BPMs around the ring. These BPMs measure proton and antiproton 53 MHz RF bunches simultaneously by having pickups at each end of the detector.

The voltage output of the detector is the sum of the signals produced at the ends and is proportional to the beam current and plate impedance,

$$\frac{V\,plate}{I\,beam} = \frac{Z0}{2}\frac{110^0}{360^0}\left[\sin\omega\left(t + \frac{\Delta t}{2}\right) - \sin\omega\left(t - \frac{\Delta t}{2}\right)\right]. \quad (2.1)$$

$$\Delta t = \frac{2l}{C_0} = 1.2 \text{ nanosec}, \quad (2.2)$$

plate of length l , $C_0$ is the speed of light.

Plate is $\frac{\lambda}{4}$ long and the peak frequency response is 407 MHz, calculated from

$$\frac{\omega \quad l}{C_0} = \frac{\pi}{2}. \quad (2.3)$$

The stripline directional-coupler design of the Tevatron BPM pickups do not offer perfect isolation between signals from particles travelling in opposite directions (P and PBars). They offer only 26 dB isolation and the coupled signal at the downstream end of the detector is 1/20$^{th}$ the amplitude of the upstream signal. With the typical 4:1 proton-to-antiproton bunch intensity ratio, this isolation alone is insufficient to support millimeter-accuracy for antiproton position measurements in the presence of protons. An accurate and manageable solution is needed when making antiproton measurements in the presence of protons that will reduce antiproton bias when making proton measurements. Two avenues of approach are suggested: first, separate the signals in the time domain, and second, calibrate the cross-talk in the frequency domain and make compensation before computing the beam position.

## 2.3 BPM System Hardware and Functionality

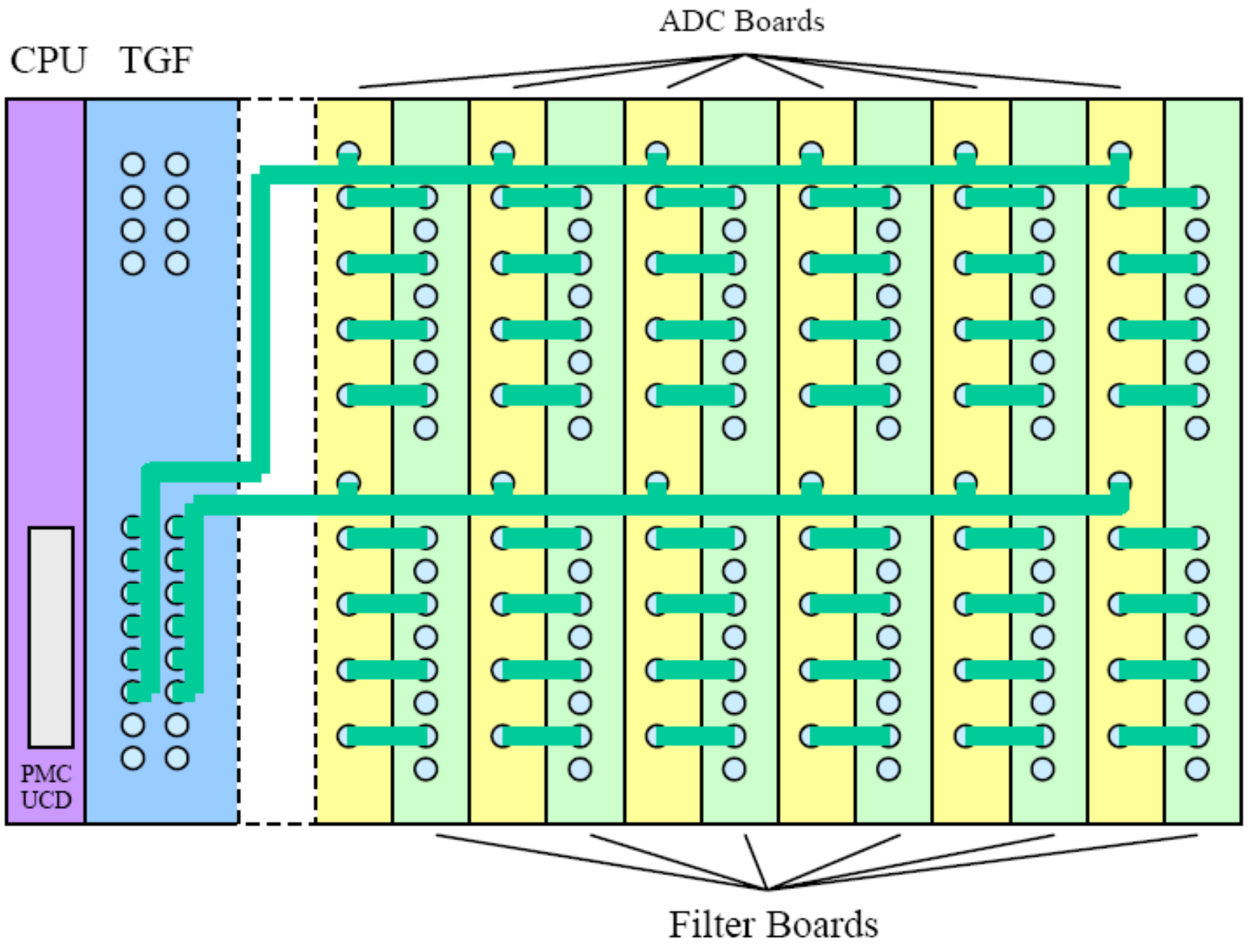

**Figure 1.** The VME64X crate contains the slot zero controller to the left (MVME 2400), followed by the timing board (TGF) which produces the triggers and clocks for each of the down-converter boards (DDC). Between each Digital down-converter board, or as shown in this figure, ADC board; there is the analog board which conditions the BPM signal from the tunnel before its digitized.

Based on experience gained in the Recycler BPM system, the upgraded Tevatron BPM readout system was designed around a VME64X crate holding a crate controller with a PMCUCD daughter board. PMCUCD stands for Universal Clock Decoder which is designed to decode Tevatron clock events and works in conjunction with the accelerator control system. A Timing Generator Fan-out (TGF) board and up to 6 digital down-converter boards (DDC) and their corresponding analog conditioning filter boards. There are 27 houses around the Tevatron ring, each having one BPM crate to handle up to 12 BPMs. Illustrated in Figure 1 is the hardware organization within a VME 64X crate. The TGF board is a re-design of the timing board used in the Recycler ring BPM upgrade, its functionalities include:

i. Phase lock to the $f_{RF}$ = 53.1 MHz (Tevatron RF frequency) and use it to generate the $\frac{7}{5} \times 53.1$ MHz = 74.3466 MHz. This is the clock frequency used by the DDC board.

ii. Decode the events transmitted through the Tevatron TCLK system to provide advanced arming signals to the crate controller (MVME 2400 processor) so that the hardware can be configured in time for different measurement types. Or present pre-defined requests for data transfers from the crate controller to the online software.

iii. Decode the events transmitted through the Tevatron Beam Sync (TVBS) system to derive accurate timing/triggering signals to synchronize up to 8 DDC boards with respect to the different beam arrivals.

iv. Generate diagnostic 53.1 MHz TTL pulses which can be continuous, or a train of seven pulses distributed across the VME backplane to feed the filter boards. The train of seven pulses is triggered by the Tevatron turn makers and adjusted in time by a varying delay from the timing board.

v. Control the switching of diagnostic signal relays on up to 8 filter boards so that the 53.1 MHz diagnostic pulses can be directed to the ADCs on the DDC boards, or back through the BPM pick-ups, or both.

The digital down-conversion boards (DDC) used in the system are model ECDF-GC814-FV-A from Echotek Corporation. The DDC board consists of eight Texas Instrument 14-bit AD converter, eight GC4016 digital down-converters, FPGA, RAM, and VME interface components [3].

The signal processing is based on an under-sampling technique. The highest sampling rate the DDC board is capable of digitizing a signal is 80.0 MHz, which does not meet the Nyquist sampling criteria ($2\times f_{max}$) required to sample the beam frequency (53.1 MHZ). By under-sampling the signal at a frequency below Nyquist (74.34 MHz), the digitizer produces a 21.24 MHz signal that is down-converted to baseband by a numerical controlled oscillator (NCO) (built-in into the GC-4016 DDC chip). After down-converting to baseband the signal is filtered in three stages. The first filter is a five stage cascaded integrated comb (CIC) filter whose Z transform response is

$$H_Z = \frac{\left(1 - z^{-D}\right)^n}{\left(1 - z^{-1}\right)^n}. \qquad (2.4)$$

D is the decimation rate of the filter, n the order of the filter. The decimation rate controls the cutoff frequency of the filter whose Z transform response is

$$H_Z = \sum_{n=0}^{N-1} h_n Z^{-n}, \qquad (2.5)$$

$$Z = r\, e^{j\omega}, \qquad (2.6)$$

r is magnitude and ω is the angle of z.

The second filter is the CFIR, which is a 21 tap Finite Impulse Response filter designed to compensate for the drooping of the CIC filter due to its SINC function response. The third filter is the PFIR, which is a 64 tap Finite Impulse Response filter and is responsible for the overall frequency response of the filtering process. There is a fourth filter implemented in the FPGA which averages the in-phase (I) and in-quadrature (Q) outputs of the PFIR filter by picking the I and Q samples whose sum intensity are above a threshold α.

**Table 3**. Frequencies of each stage of the GC-4016 Digital Down-Converter IC.

| | |
|---|---|
| Beam RF Frequency (Hz) | 53.1048e6 |
| ADC Sampling Frequency (Hz) | 74.34672e6 |
| NCO Frequency (Hz) | 21.24192e6 |
| CIC Filter Clock Frequency (Hz)<br>Decimate ADC from 4 to 32565 | 18.58668e6 (when fclock/4) |
| CFIR filter clock frequency (Hz)<br>Decimate CIC filter clock by 2 | 9.29334e6 (when CIC decimation = 4) |
| PFIR filter clock frequency (Hz)<br>Decimate CFIR filter clock by 2 | 4.64667e6 (when CIC decimation = 4) |

Channels 1 and 2 of the DDC board digitize the signals from the upstream ports on the proton side of the BPM (commonly referred as pickups A and B).

Channels 3 and 4 digitize the signals from the same BPM but from the down-stream pick-ups on the antiproton side. Channels 4 through 8 repeat this pattern.

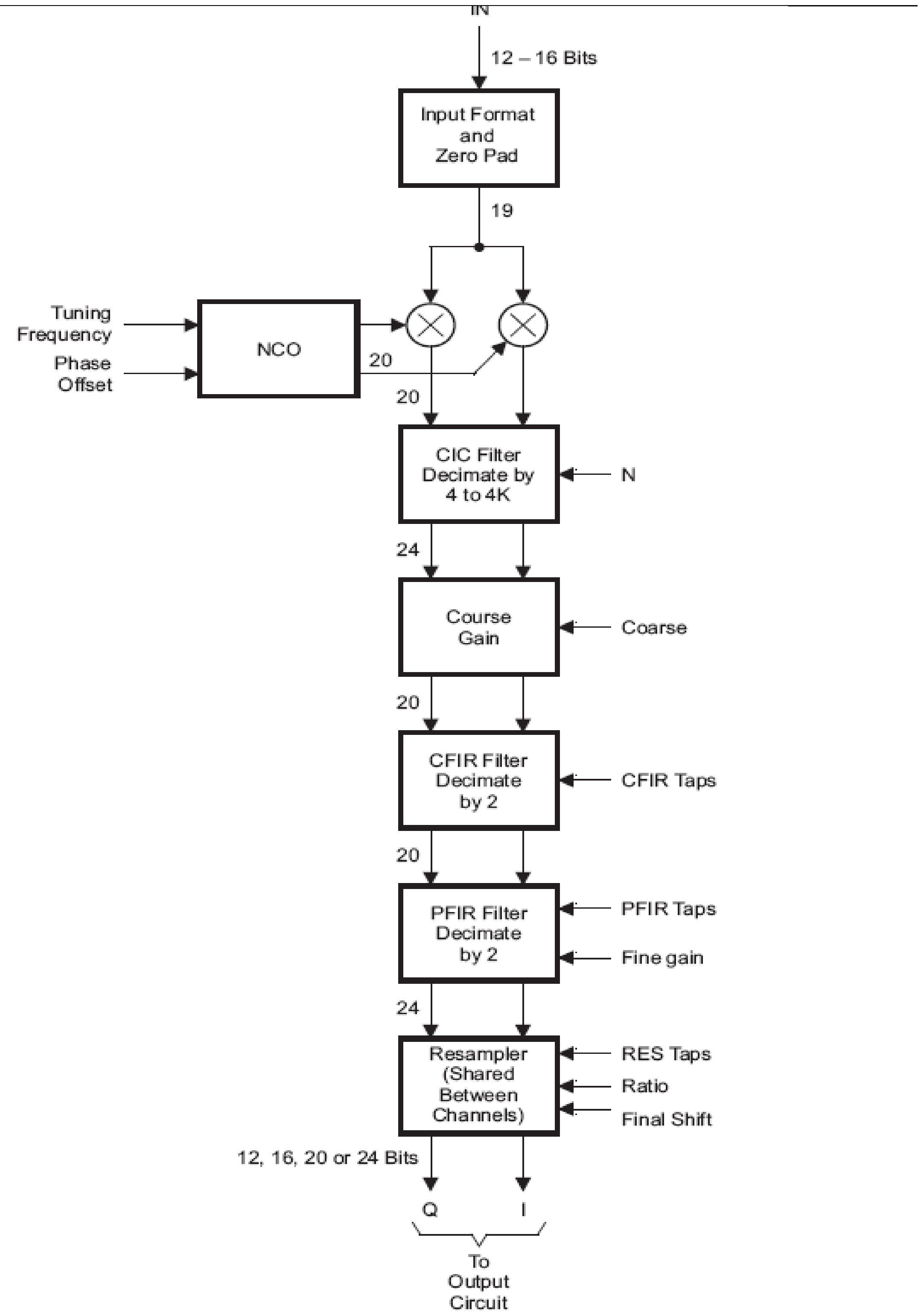

**Figure 2.** Block diagram of one-of-four signal processing paths in the GC-4016 digital-down converter IC. There are eight GC-4016 devices on the DDC board and each has four signal processing devices shown in this figure. Each ADC output provides an input to each down-

converter signal processing path allowing four parallel, independent signal processing processes to take place simultaneously [3].

### 2.4 Tevatron Events Used for Arming and Triggering

**Table 4.** TCLK Timing Events in the Tevatron.

| Event | Description | Comment |
|---|---|---|
| $71 TCLK | Tev:BPM Prepare for beam. | Presently issued 0.01 secs after $4D. Will be issued once before shot setup. |
| $73 TCLK | Tev:BPM High field. | Issued half way up the energy ramp. Was used to change alarm limits for BPMs. Not needed for the new system. |
| $74 TCLK | Tev:BPM Low field. | Issued after a $4D. Was used to change alarm limits for BPMs.<br>Not needed for new system. |
| $75 TCLK | Tev:BPM Write profile memory. | Trigger times programmed into a CAMAC 070 Card. Used to collect profiles up the ramp. |
| $C1 TCLK | Tev:Tevatron reset for collider operations. | Enabled by sequencer and triggered by $41. Used to reset display frame pointer. |
| $C2 TCLK | Tev:Prepare to accelerate collider beams. | Referenced to $41 event. Used to reset the profile frame pointer. |
| $78 TCLK | Tev:BPM Write display frame. | Variable time and trigger. Typically set to 2.71 secs after a $4D. |
| $40 TCLK | Tev:Reset for pbar injection @150GeV. | Start of Pbar injection from MI->Tev. Occurs about 2.7 seconds before the actual injection. |
| $5B TCLK | Collider pbar beam transfer trigger from MI to Tevatron. | TCLK echo of MIBS $7B. MI->Tev transfer of pbars happens on the MIBS $7B which is about 2.7 seconds after the $40. |
| $4D TCLK | Tev:Reset proton injection @150Gev. | Start of proton injection from MI->Tev. Occurs about 2.7 seconds before the actual injection. |
| $5C TCLK | Collider proton beam transfer trigger from MI to Tevatron. | TCLK echo of MIBS $7C. MI->Tev transfer or protons happens on the MIBS $7C which is about 2.7 seconds after the $4D. |
| $47 TCLK | Tev:Beam has been aborted. | TCLK event generated when the BSSB pulls the Tevatron abort. |
| $4B TCLK | Tev:Abort clean up. | TCLK event used to trigger the Tev abort kickers and intentionally remove beam.<br>(Every time beam is removed from the Tevatron either a $47 or $4B is issued.) |

**Table 5.** Beam Sync events of the Tevatron.

| Event | Description | Comment |
|---|---|---|
| $AA TVBS | TeV: Beam | Tevatron Beam Sync event turn marker |
| $ 7C TVBS | Main Injector : Ini MI -> TeV proton collider | Tevatron Beam Sync event at injection derived from Main Injector $7C |
| $ DA TVBS | Trigger TBT data collection | Tevatron Beam Sync used to trigger TBT data collection. Synchronize all BPM's to same turn. |

### 2.5 Timing Modes

Timing signals from the data acquisition system operate in one of two basic modes: repetitive and one-shot. The former is used for closed orbit measurements, while the later is used for the various types of turn-by-turn (TBT) measurements.

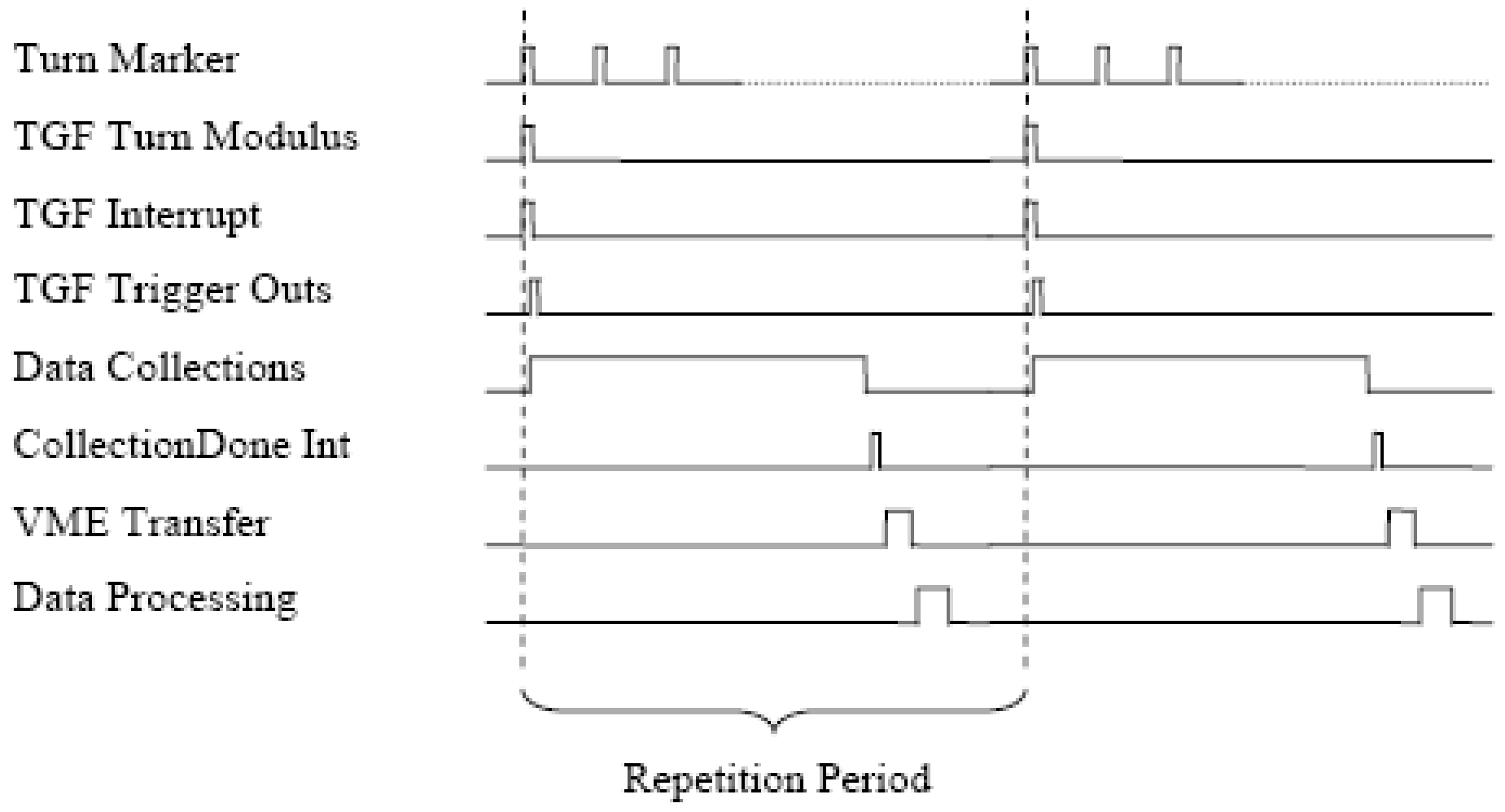

**Figure 3.** Timing during closed orbit mode.

For closed-orbit measurements, the front-ends in each house are not synchronized to the same turn of the corresponding bunch, they only cover the same length in time. In this mode, the TGF simply decimates the Tevatron revolution/turn markers, generates an interrupt to inform the crate controller of a pending data acquisition, applying appropriate time delays to account for the different bunch arrivals at different BPM pick-ups. This allows data from all the BPMs to have the same phase, and generates trigger (SYNC) signals for the DDC boards. When the pre-set number of samples (burst counts) has been collected, the first DDC board in the crate issues a collection-done interrupt then notifies the crate controller that a data transfer will take place. The timing dependencies are plotted in Figure 3.

For Turn-By-Turn (TBT) measurements, the front-ends in all the houses are synchronized to the same specific turn, so that the measurements they make correspond to the same beam. This is achieved by setting the TGF to wait for a particular “start-event”. Upon the occurrence of this “start-event”, the TGF generates an interrupt to inform the crate controller of a pending data acquisition, then repeats the following steps for the desired number of turns: wait for the next “turn-event”, be it the turn markers, apply appropriate time delays and generate trigger signals for the DDC boards. When the pre-set number of triggers has been received, the first DDC board in the crate issues a trigger-counter interrupt. These timing relationships can be seen in Figure 4.

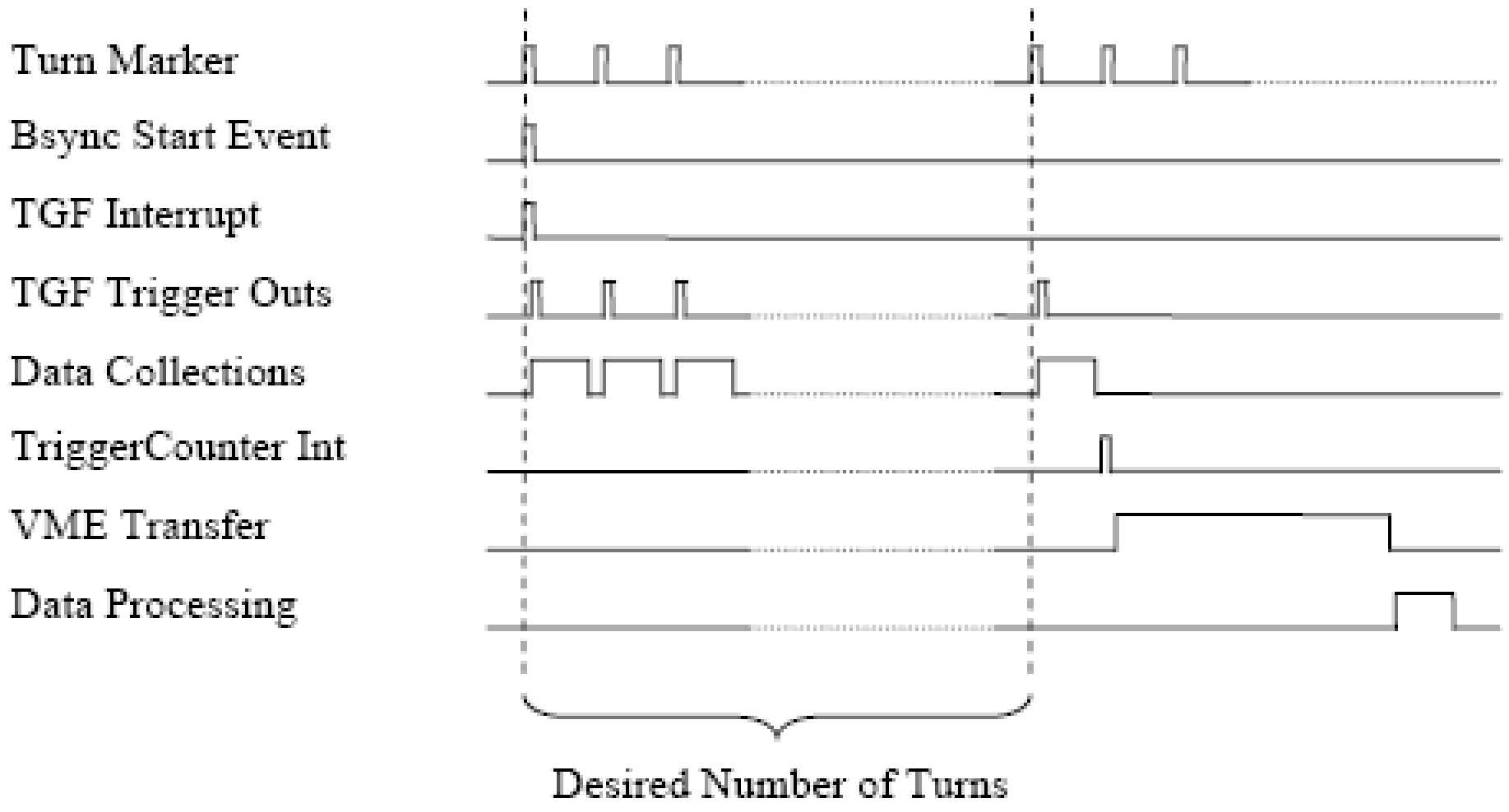

**Figure 4**. Timing during Turn-By-Turn mode.

It should be pointed out that for a given state of the Tevatron, the time delay from the Tevatron turn marker (TVBS $AA) as seen by the TGF until the actual bunch passing is fixed at each BPM location. In order to ensure that the ADCs in the DDC boards are triggered at the same time relative to the actual bunch passing of the same turn, 4 layers of delays are implemented. These delays are global delay, house delay, DDC board delay and channel pair delay. In different Tevatron states especially if the beam is injected to different buckets, these delays may need to be adjusted if the same timing relationship is required [4].

**Table 6.** Timing Delays needed to synchronize beam and triggers in the Tevatron.

| TeV Service Building | BPM | Pre-Trigger Delay (Turns, 1 turn=20.95 uSec) | House Delay (53MHz Buckets) | Board delay (53MHz Buckets) | Channel Delay (7/5 * 53MHz Buckets) |
|---|---|---|---|---|---|
| A1 | HPA11 | 6 | 970 | 31 | 0 |
| | HAA11 | | | 31 | 0 |
| | VPA11 | | | 31 | 0 |
| | VAA12 | | | 31 | 0 |
| | VPA12 | | | 23 | 0 |
| | HPA12 | | | 23 | 0 |
| | HPA13 | | | 23 | 0 |
| | HAA13 | | | 23 | 0 |
| | VPA14 | | | 12 | 9 |
| | VAA14 | | | 12 | 9 |
| | HPA15 | | | 12 | 7 |
| | HAA15 | | | 12 | 7 |
| | VPA16 | | | 40 | 7 |
| | VAA16 | | | 40 | 7 |
| | HPA17 | | | 40 | 16 |
| | HAA17 | | | 40 | 16 |
| | VPA18 | | | 80 | 7 |
| | VAA18 | | | 80 | 7 |
| | HPA19 | | | 80 | 21 |
| | HAA19 | | | 80 | 21 |

This table is an example of the four timing parameters required to synchronize beam arrival at each BPM with the trigger to that DDC board in one house. An RF bucket refers to the period of one 53.1048 MHz cycle.

The house delay is the time-of-flight of the beam around the ring from the point of injection up to a particular house and the arrival of the TCLK event triggering coinciding with the same BPM house. The cable delay between the BPM and the ADC is measured for each pair of BPM pickups making up the board delay channel delay. These delays are referenced to the injection kicker timing and the location of the building to the proton injection location which is F0. These times have not been changed since the BPM system was commissioned due to stability of the timing distribution around the ring.

## 2.6 Data Acquisition Modes

The BPM data acquisition system has several operation modes, they are mutually exclusive due to the specific configurations of triggering and filtering in the DDC boards.

Idle mode is the "no-function" mode of the BPM operation that occurs when there is no beam in the Tevatron. During this mode, all operational data buffers are frozen, and the TGF is set to wait for TCLK $4D. Upon the detection of TCLK $4D, the TGF interrupts the crate controller and the system transits into the First Turn mode.

Closed Orbit Mode. This is the default mode of operation when there is beam in the Tevatron. In closed orbit mode the DDC boards are set up so that one signal channel in each GC4016 signal processing IC performs the in-phase (I) calculations and a second channel performs the in-quadrature (Q) calculations. This reduces the data decimation factor by two and generates an output at every clock cycle. The CIC filters decimate by 1024 bringing the total decimation rate to 4096, the overall effect on the output data rate is that its reduced by this factor. This corresponds to a PFIR output data rate of about 18 kHz while the re-sampler is bypassed. Both CFIR and PFIR filters implement a boxcar filter, of 9 and 23 taps respectively. Combining the CIC, CFIR and PFIR, the overall 3 dB bandwidth is about 700 Hz which is dominated by the PFIR coefficients.

The narrowband measurements are triggered at 500 Hz. The triggers are sourced by the TGF board through a decimate-by-N counter by down-sampling the 47 kHz Tevatron turn marker (TVBS $AA). The value of N, and thus the trigger rate, can be adjusted through the turn modulus/decimation register on the TGF board. At 500 Hz, the data acquisition system has about 200 µs idle time between triggers.

Turn-By-Turn Mode. During turn-by-turn (TBT) operation, the BPM system samples the position of the beam at every BPM, at every turn for 8192 turns. The primary use of TBT measurement is at bunch injection. The selection number of turns was chosen because it provided a good view of the damping of injection oscillations and provided sufficient data to compute the machine tune from these oscillations. Data is analyzed to determine Tevatron lattice information such as beta functions, phase advance, local coupling.

In this mode, the DDC signal processing channels operate in the split I-Q fashion. The total data decimation is reduced to 16 from 32. The corresponding data output rate is about 4.65 MHz. Both CFIR and PFIR filter coefficients implement a boxcar filter consisting of 13 taps in the PFIR case and 2 taps in the CFIR case. The combined overall bandwidth is about 360 kHz which is dominated by the PFIR filter frequency response.

On successful completion of a turn-by-turn cycle, the data is transferred to the crate controller using the direct memory access mode (DMA) by the DDC boards. The positions and intensities of the proton bunch at each BPM location are calculated in the same fashion as in the closed orbit mode. The data structures are then stored in 8192 point deep TBT buffers for later access. After a TBT measurement, the DA system switches to closed orbit mode as soon as possible. Due to the large data volume there is a dead time of a few milliseconds.  The cycle time available to the BPM system is separated into:

i. Start of cycle mode – There are at least 50ms between the TCLK reset event and the arrival of beam. This is to allow the front end to configure the timing boards and DDC boards.
ii. Mode Change – The front end must be able to change modes in less than 10ms, i.e. - going from Closed Orbit to Turn By Turn mode.

iii. End of Cycle Mode– There is at least 500ms between the end of beam signal and the beginning of the next cycle in order for the front end to read data off of the DDC boards.

After the DDC boards have completed data acquisition, the crate controller reads out, through VME single reads, one *I-Q* pair from each channel, four pairs for each BPM. The proton contaminations on the antiproton signal are removed from the raw antiproton I's and Q's (A and B plates) according to:

$$I'_{Pbar_r} = I_{Pbar} - C_I \times I_P + C_Q \times Q_P \ . \tag{2.7}$$

$$Q'_{Pbar} = Q_{PBar} - C_I \times Q_P + C_Q \times I_P \ . \tag{2.8}$$

where the subscripts P and Pbar denote proton and antiproton signals, and $C_I$ and $C_Q$ are the in-phase and quadrature components of the coupling coefficient. The modulus of each channel (*M*) is then calculated:

$$M = G \times \left( \sqrt{I^2 + Q^2} \right) - M_0 \ , \tag{2.9}$$

where G and $M_0$ are gain and offset of the electronics. Finally, the beam positions $D_p, D_{pbar}$ and the intensities $S_p, S_{pbar}$ can be determined according to:

$$D = G \times \frac{(M_A - M_B)}{(M_A + M_B)} - D_M \ . \tag{2.10}$$

$$S = M_A + M_B \ . \tag{2.11}$$

where *G* is a scale factor to convert the unit-less quantity to millimeters (nominally 26 mm), $D_M$ is the mechanical offset that was surveyed relative to the BPM's electrical center before the BPM was installed in the ring. For each BPM, the 2 positions $D_{P,Pbar}$ and 2 intensities $S_{P,Pbar}$ are put into a data structure together with the 4 raw *I* and *Q* pairs (see data structure section). This data structure is then stored in a 1024 point deep circular buffer, the Fast Abort buffer, and propagates to the other buffers.

## 2.7 Data Structures

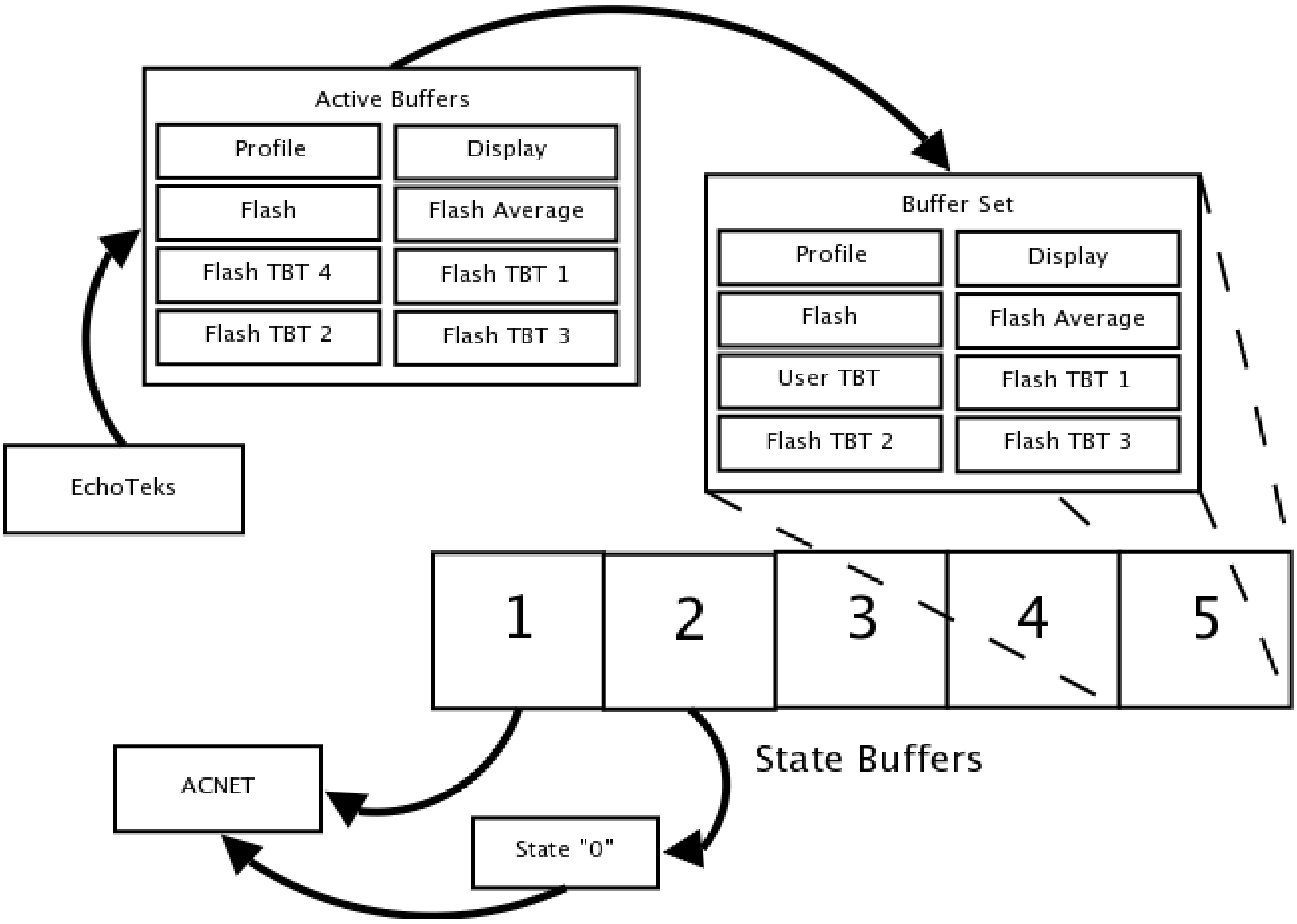

**Figure 5**. Data buffer structure in the BPM system. The DDC stores data according to the measurement type. Once the measurements are completed, the buffers are transferred to the crate controller and stored related to the machine state.

In closed orbit mode, the data acquisition software maintains the following data buffers:

a. Fast Abort buffer – circular buffer, 1024 points deep. One element in the fast abort buffer is referred to as a frame**.** The crate controller takes measurements from all BPMs every 2 milliseconds, and puts them into the fast abort buffer. The process stops on Tevatron aborts (TCLK $47 or $4B), and re-starts automatically after the Injection mode.
b. Slow Abort buffer – circular buffer, 1024 points deep. Data is copied from the newest frame in the fast abort buffer every 500 readings where the value 500 is configurable.
c. Profile Frame buffer – FIFO, 128 points deep. Data is copied from the latest frame in the fast abort buffer every time a TCLK $75 (shot data event) is received, i.e. each index in the profile frame corresponds to a particular shot data event. If the profile frame buffer overflows, new values are discarded and an alarm condition is flagged -- only one alarm is sent per crate. If a $75 event arrives when in a mode other than Closed Orbit, this particular profile frame is filled with a bad status (999).
d. Display Frame buffer – single frame. Data is copied from the latest frame in the fast abort buffer every time a TCLK $78 (display event) is received.

e. Fast Time Plot (FTP) buffer(s) – data is copied from the latest frame in the fast abort buffer every 2 milliseconds. There is no depth to this buffer, but since fast time devices plot only single values, it will be a series of devices to plot:

    i. A particular BPM's proton position.
    ii. A particular BPM's antiproton position.
    iii. A particular BPM's proton intensity.
    iv. A particular BPM's antiproton intensity.
    v. A particular input's I value.
    vi. A particular input's Q value.

f. Sum signal for each channel (magnitude of A + magnitude of B).
g. Snapshot Data request – the most recent frame in the fast abort buffer is copied to the request's buffer.

When a TCLK $47 (abort) or TCLK $4B (no beam left in machine) is received, the DA system does the following:

Freeze data in the slow abort buffers, the profile frame buffer, and the display frame buffer.

a. Fill the FTP buffers with a status indicating no beam.
b. Acquire a few more measurements and fill the fast abort buffer, then freeze it.
c. Change the mode of operation to the idle mode.

### 2.8 Operational Modes

First Turn Mode. The first turn mode is a special case of the turn-by-turn mode and is intended to capture the position of the beam at each BPM location as the beam travels its first revolution around the ring immediately after injection, while the injection bumps are still active. This position information is used to diagnose errors in injection position and angle in order to tune the magnets in the injection beam lines.

Since the injection bump magnets are only activated briefly, it is not feasible to take a single turn TBT measurement, then reconfigure the electronics and switch to closed orbit mode to catch the injection closed orbit. Instead, the DA system will go into a normal 8192-turn TBT measurement, retrieve the first turn position, and obtain the injection closed orbit by simply averaging over all elements in the injection TBT buffer that have intensity above a predefined threshold.

The first turn mode is armed if the TGF receives a proton injection event (TCLK $4D) and the Tevatron state device indicates "no beam". The start event for the TGF is TVBS $7C which corresponds to the kickers being fired during Main Injector-to-Tevatron transfers. The TCLK $4D occurs about 2.5 seconds before TVBS $7C, allowing ample time for the DA system to configure the electronics. The first turn measurement will be stored in the First Turn buffer for later readout.

Asynchronous Injection Mode. In this mode the DDC boards are configured to take 8192 raw data samples for each channel at a sampling frequency of 74.3466 MHz, these samples cover about 70 revolutions.

The TGF is set to trigger at the first Tevatron turn marker after a specific start event. The global (coarse) delay is used to make sure the 70 revolution window covers the first turn. Fine (short time) delays like house delays, DDC board delays and DDC channel pair delays can be used but not required.

The raw data can be used online to determine the aggregate time delays for each BPM location through certain pattern recognition. It can also be used offline to calculate the beam positions and intensities so as to help diagnosing the injection beam lines.

Calibration Mode. The calibration mode is basically a normal run mode where data collected is tagged as calibration data. This specially marked data is used on the offline processing to define the calibration constants to be used by the front-end when calculating the position of the beam.

The user sets the calibration mode and passed on to the front-end DAQ system through ACNET variables. All data read out from DDC cards are tagged as calibration data, until the calibration mode is disabled.

## 2.9 Diagnostics

The diagnostic application program has the ability to control the filter board switches individually. To simplify the configuration to the user, there are four possible relay configurations: diagnostic off, diagnostic to DDC board only, diagnostic to tunnel only, diagnostic to DDC board and tunnel. The application allows the tests above to be generated automatically for a single house upon a user request.

These measurements are compared to baseline measurements for possible variations that are out of tolerance. Those measurements that are out of tolerance by more than 1% in magnitude are flagged as potentially requiring a new calibration. The user also has the option of saving the results as the new baseline.

The next level of diagnostics for the system is monitoring and control of the VME crate. The RS232 interface on the crate is used to control and monitor the voltage levels and fan speeds on each of the crates. An Ethernet to RS232 adapter is utilized as the communication bridge to the internal crate monitoring and control. The adapter is powered independently from the crate to allow a remote user to power down the crate without losing communication to the adapter. The remote control of the crates includes the ability to perform a SYSRST on the backplane, a SYSFAIL on the backplane, and power down the crate.

A third level of diagnostics of the system verifies the proper events and triggers are present. There are four types of events that are monitored for diagnostic purposes: State transitions, TCLK events, The TVBS $AA marker, Other TVBS events.

The system needs a means of monitoring the order of critical events that the timing card and processor will react to. The timing card keeps a circular buffer of the last 256 events (TCLK, TVBS $7C & $DA, measurement complete, etc.) that it uses to generate interrupts or triggers, excluding TVBS $AA events. The timing card does not record all TCLK events, only the events that it uses to generate interrupts to the processor. The processor, in turn, has a circular buffer of the last 256 events of interrupts that it receives from the timing card as well as any other events that cause a change in system configuration (change in state device V:TEVBPM, request for change into diagnostic mode, etc. These two buffers are accessible from the Tevatron BPM diagnostics application and are used to verify that the processor is receiving the proper interrupts and that the events are happening in the proper order.

An event diagnostic that is needed which detects the presence of the $AA marker. This is performed on every turn-by-turn and first-turn data acquisition. While triggering the turn-by-turn measurement, the timing card verifies that the $AA markers are being triggered every turn by comparing the time between markers with a divide by 1113 counter clocked by the 53.1 MHz. If a trigger is missed, the turn number that has been skipped is loaded into a buffer. The data user can then reconstruct the proper sequence of data for analysis.

The fourth level of diagnostics is verification of the DDC board and GC4016 setup. The diagnostic application is able to override the default DDC board configurations and force it into closed orbit mode, turn-by-turn mode, or raw ADC mode. There are means to override the trigger card configuration as well, so that the system can be configured to trigger the DDC board on after an arbitrary TCLK event and delay.

The software retrieves data from all active channels of the GC4016 (4 channels) chip. It does this for two of the eight DDC board channels. The data from separate GC4016 channels is organized and plotted after each trigger. The application has the means for saving channel data for offline analysis, either by saving to a file, or e-mailing the data to a users e-mail account.

The final level of diagnostics is the ability to halt and examine all of the processor data buffers. There is an option on the diagnostic application page to halt data acquisition into the buffers for the DDC channels selected (up to two). The application allows comparisons between the processor buffers and any associated ACNET variables that mirror the buffers, discrepancies are flagged.

## 3. Discussion

### 3.1 Systematics

The accuracy and resolution of the BPM system ultimately depends on factors such as cable impedance, voltage standing wave ratio (VSWR), attenuation, reflections, detector impedance, directionality, receiver noise, frequency dependent linearity, frequency dependent errors producing position time errors. All of these factors are to some extent understood, but further measurements need to be made with the BPM system to sort out the contribution of each of them to the intensity and position resolution.

## 3.2 TBT Results

Turn-by-turn measurements showed that synchrotron and Betatron oscillations can be measured with the new BPM system. Also effects like high voltage coupling to the beam causing position errors can be seen in the measurements [5].

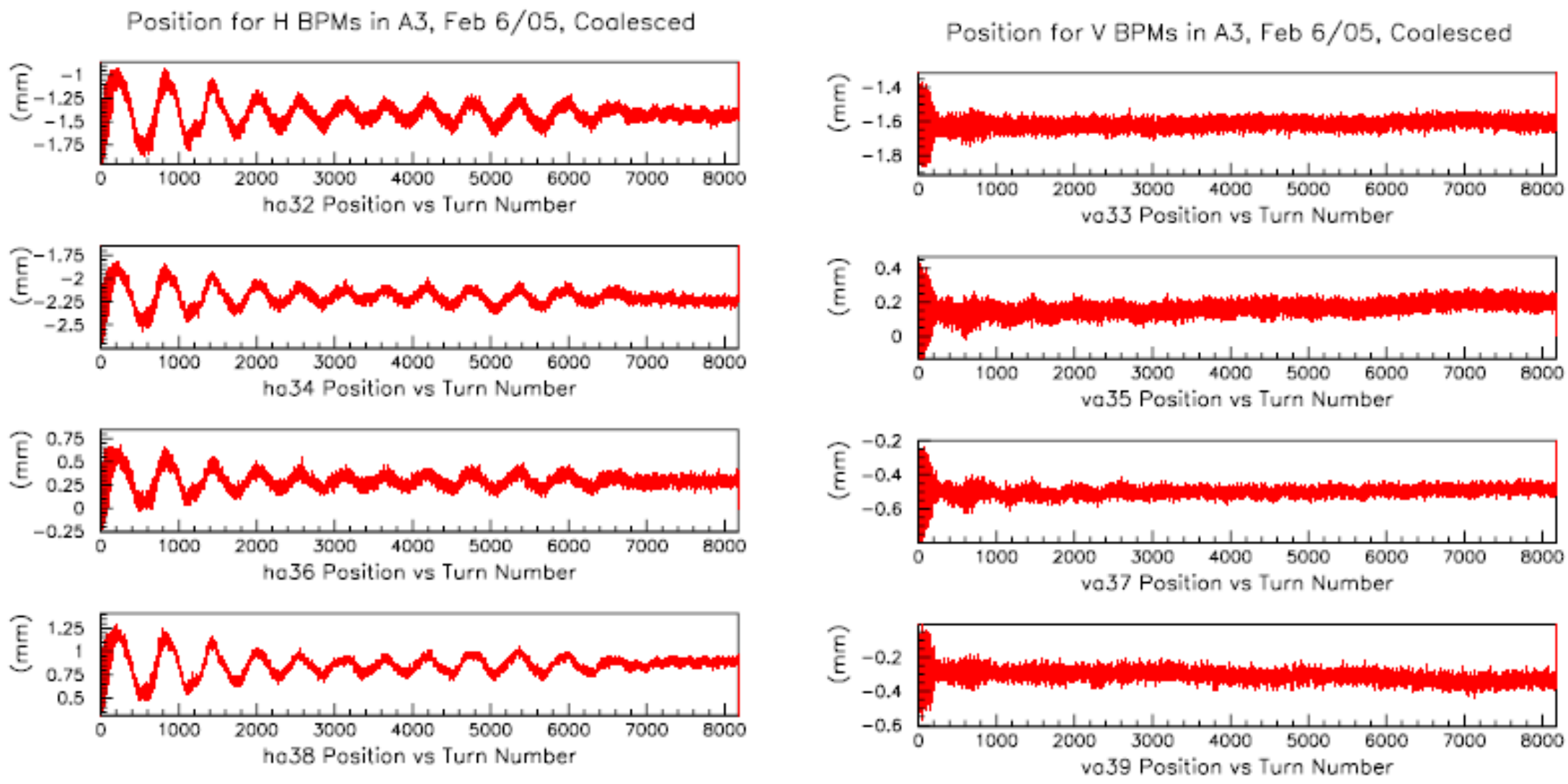

**Figure 6.** Coalesced bunch injection TBT. Figure on the left shows the oscillations in the horizontal plane of four BPMs HA32, HA34, HA36, HA38. Figure on the right shows the position during the same injection on the vertical plane for BPMs VA33, VA 35, VA37, VA39.

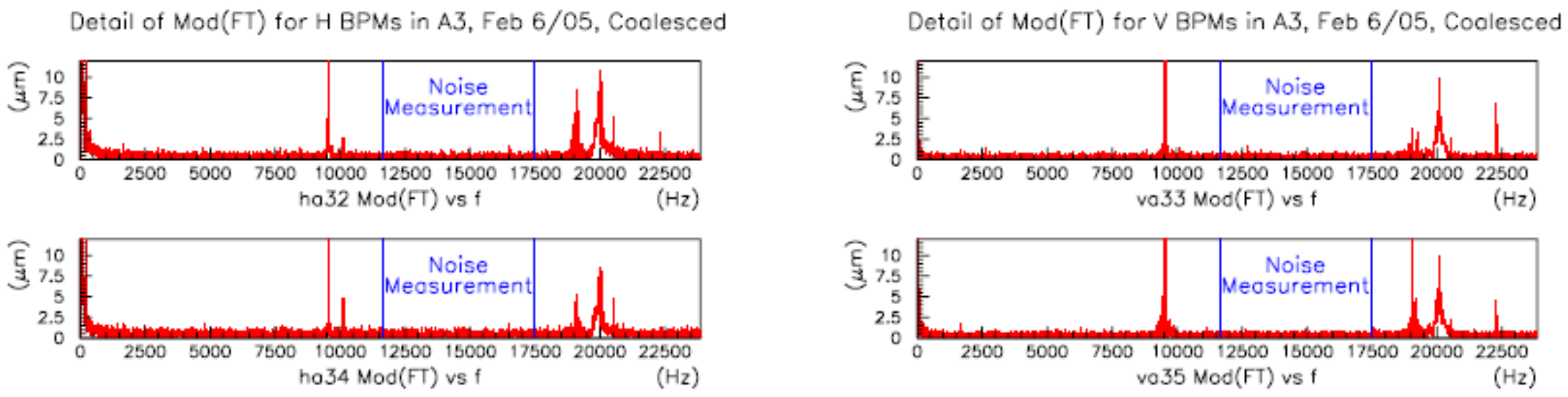

**Figure 7.** FFT of the coalesced injections showing the synchrotron line at 80 Hz, the Betatron lines at 22 KHz and artifact lines near the betatron lines and at 9 KHz. The artifact lines are 1113/5.

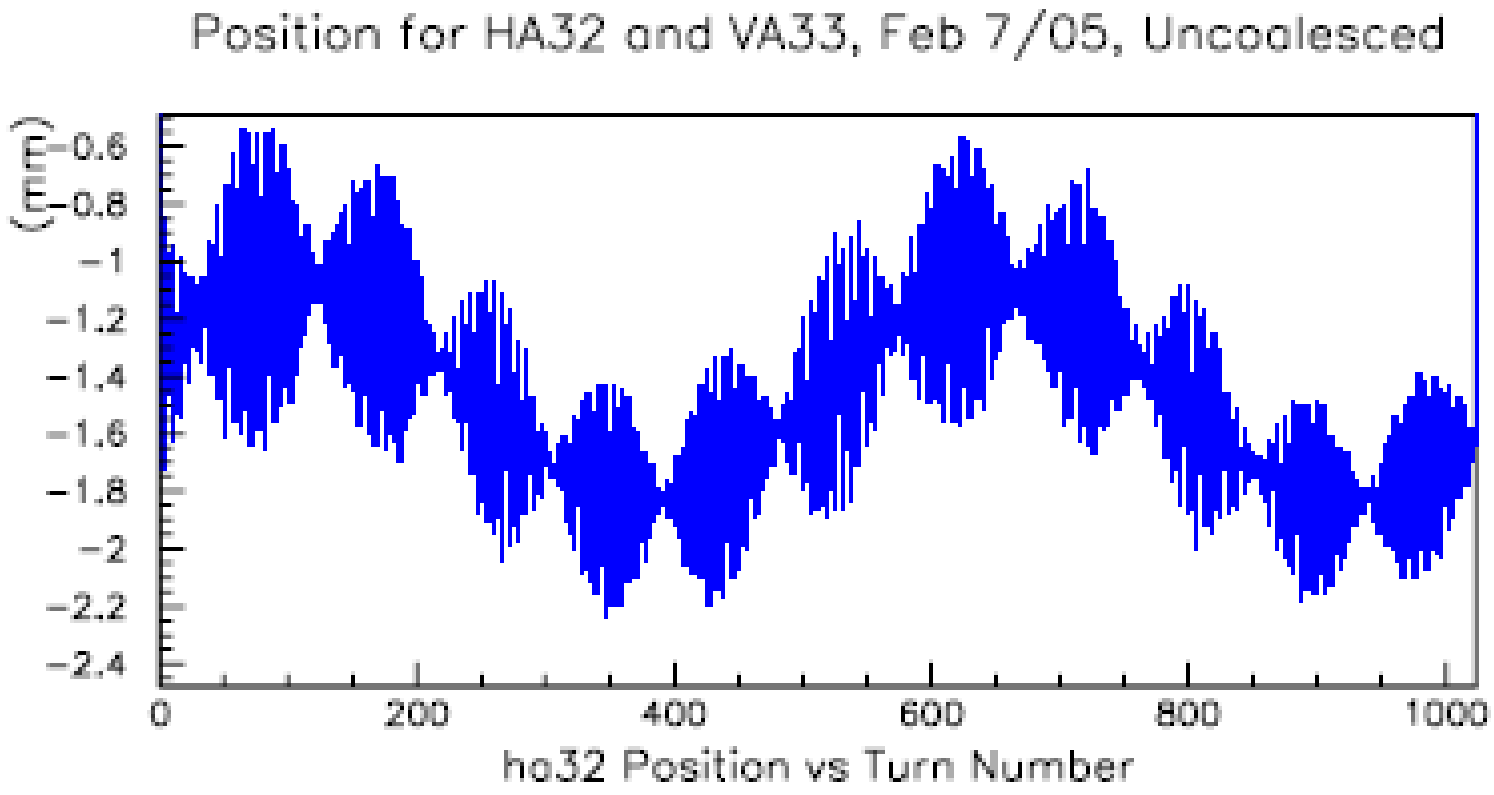

**Figure 8.** Uncoalesced injection showing high voltage coupling in the horizontal position.

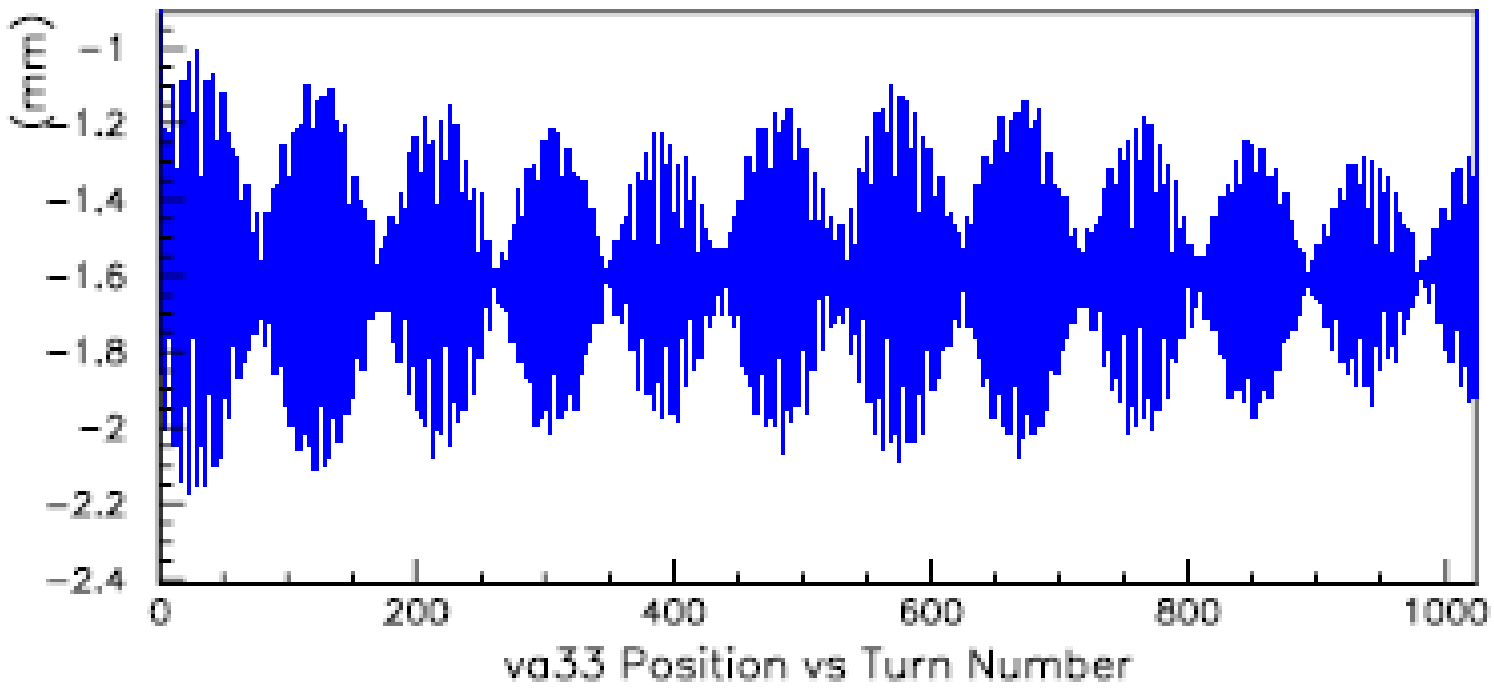

**Figure 9.** Uncoalesced injection shows no high voltage coupling.

### 3.3 Measurement Stability

The requirements document for the Tevatron BPM Upgrade [1] states that the required long term stability of the BPM system should be < 20µm. One of the challenges when assessing the long term stability of the upgraded Tevatron BPMs is how to find a reference quantity that is sufficiently stable. Beams-doc-1925 [6] shows that the measured size of the helix is remarkably stable, of order 40 µm for protons and 50 µm for anti-protons for the six weeks included in that study. This stability was achieved even though the central orbit changed by 1 or 2 mm over the course of that study (Beams-doc-1925, Figure 7) [6]. The measured size of the helix includes contributions from both true beam motion and instrumental effects. Therefore the quoted number represents an upper bound on the stability of the instrument. While this does not prove the requirement of < 20 µm is met, it is the best demonstration yet for the long term stability of the system.

### 3.4 PBar Turn-By-Turn Measurement Study

On August 29th 2011, C2 house BPM timing was configured to measure antiproton positions at injection with the new BPM system. This was tried only at one location in the ring and nowhere else. This measurement required calculating the proper house timing delays as well as the board delays, and turning off the timing parameters used for proton measurements. The first step was to measure proton injections as shown in figure 10. Then antiproton bunch was measured after cogging at the same location. The bunch intensity was 1.5e11 antiprotons. The only conclusion that can be drawn from this study is that antiprotons can be measured with the new BPM system.

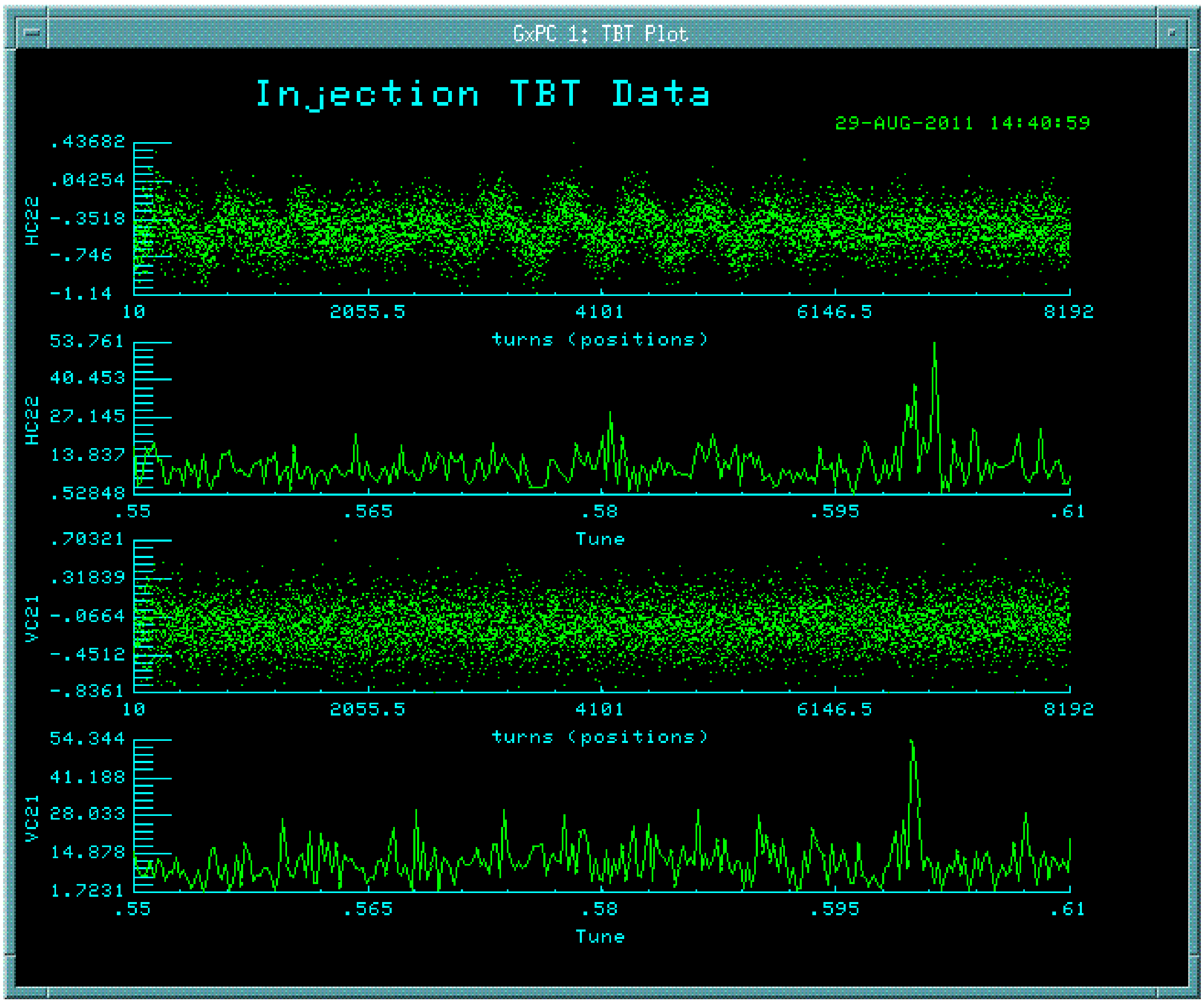

**Figure 10.** Proton injection of 8192 turns measured using horizontal BPM HC22 and vertical BPM VC21. The first 10 turns are not shown since the house is triggered 10 turns before the arrival of the beam making identification of 1st turn clear. The traces below each turn-by-turn position is the FFT of the positions and the machine tune line is calculated with this application.

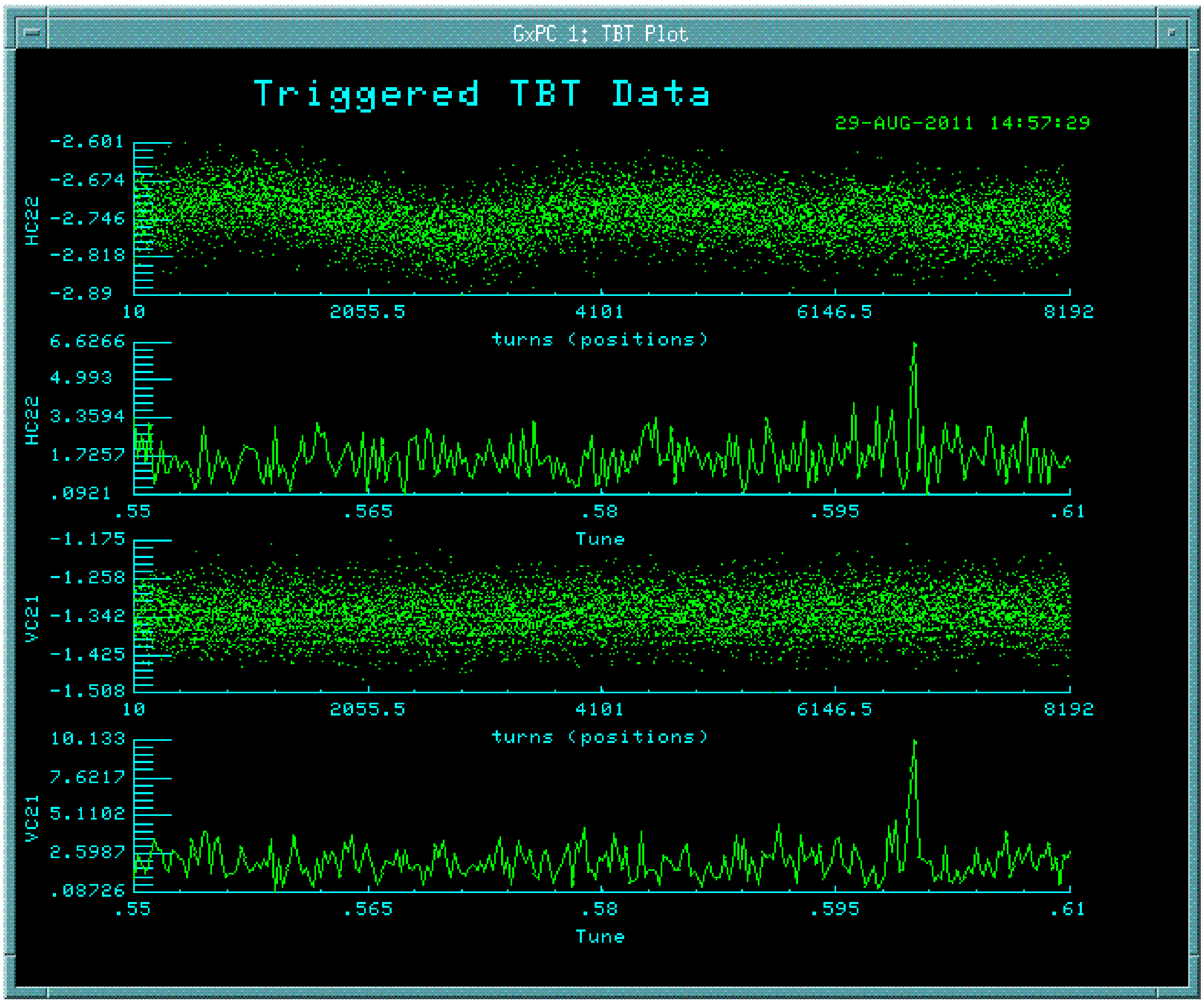

**Figure 11**. Antiproton injection after cogging measured at C2 with horizontal HC22 and vertical BPM VC21.